\documentclass[preprint,amsmath,showpacs,amssymb]{revtex4}
\usepackage[dvips]{graphicx}
\usepackage{amsmath,amssymb} 
\begin{document} 
\title{The lightest Higgs boson production at photon colliders in the 2HDM-III} 
\author{R. Mart\' inez}
\affiliation{Departamento de F\' isica, Universidad Nacional de Colombia, Bogota, Colombia}
\author{ D. A. Milan\'es}
\affiliation{Departamento de F\' isica, Universidad Nacional de Colombia, Bogota, Colombia}
\affiliation{Departament de F\' isica Te\'orica, Universitat de Valencia, E-46100, Burjassot, Valencia, Spain}
\author{J.-Alexis Rodr\' iguez} 
\affiliation{Departamento de F\' isica, Universidad Nacional de Colombia, Bogota, Colombia}
\date{}
\begin{abstract}
The branching ratios of the lightest $CP$-even Higgs boson $h^0$, in the framework of the General Two Higgs Doublet Model are calculated.  Different scenarios are presented taking into account constraints obtained in previous works on the flavor changing neutral currents  factors.  Plausible scenarios with flavour changing processes at tree level like $b \bar s$ and $t \bar c$ are analyzed for relevant region of parameters. The loop-induced Higgs couplings to photon pairs can be tested with a photon collider. The number of events of $h^0$ as a resonance in photon colliders are calculated taking into account its corresponding background signal in TESLA, CLIC and NLC.
\end{abstract}
\pacs{12.60.Fr, 14.80Cp, 13.90.+i, 13.85.Rm}
\maketitle
 
\section{Introduction} 
 
Although the Electroweak Standard Model (SM) \cite{SM} has been 
experimentally tested with excellent results, the scalar sector of 
the model responsible for the electroweak symmetry breaking 
remains without any experimental test; thus the big mismatch of 
                the SM is the absence of the Higgs boson. A wide variety of models 
have been introduced to address the puzzles of the the electroweak 
symmetry breaking and the hierarchy mass problem. For instance 
Supersymmetry\cite{susy}, large extra-dimensions\cite{extra}, 
strong dynamics leading to a composite Higgs boson\cite{strong} 
and little Higgs models\cite{little}. There are many other 
problems that are not addressed by the SM and suggest new physics 
beyond the SM; one of them is the mixing between families, which 
is related with the neutrino oscillation and the flavor changing 
neutral current (FCNC) processes\cite{osc}. 
 
One simple extension of the SM is the so called two Higgs Doublet 
Model (2HDM) which add a new Higgs boson doublet with the same 
quantum numbers as the fist one, and in this way allow a new 
rich variety of phenomenology. In one attempt to solve the hierarchy problem for the quarks in the third family, one kind of these 2HDMs give masses to the up sector with one Higgs doublet and to the down sector with the other one.
This model is the so-called 2HDM type II, and precisely it is the one that 
 appear in the framework of the minimal supersymmetric 
standard model \cite{susy}. The FCNC processes can be avoided at 
tree level in the 2HDM by imposing discrete symmetries \cite{GW}. 
This symmetry leads to 2HDM type I or II. However,  our study 
is going to focus on the 2HDM type III without any discrete 
symmetry, the most general version of the 2HDM, in which the FCNC 
processes appear at tree level in the Yukawa couplings. 
 
From the experimental data at LEP collider  has been found the 
bound $m_h \geq 114$ GeV, through direct search of the Higgs 
boson\cite{hep-ex0412015}; On the other hand, radiative 
corrections of the electroweak parameters lead to a light Higgs boson 
in the SM scenario with mass below 220 GeV\cite{precision}. 
Otherwise, TeV scale is being tested by the Fermilab Tevatron and 
will soon be explored at the CERN Large Hadron Collider (LHC). If the Higgs boson is SM like its discovery is guaranteed at the LHC; its mass will be measured with high precision and in some channels, the signals will be enough to allow the knowledge of certain combinations of Higgs partial widths up to $10-30 \%$ level \cite{LHC}.  Further, the linear 
$e^+ e^-$ collider will study the dynamics of the new physics with 
high precision\cite{linear}; it will measure the production cross section of a light Higgs boson in Higgsstrahlung or $WW$ fusion as well as the important branching fractions with a few percent level. 
 
It is a challenge the discovery of the Higgs boson; but an 
important issue is to decide what kind of model it is coming from, 
which can be elucidated through its coupling with standard particles. In the 
standard model their couplings are always proportional to the 
particle masses. Other models have deviations \cite{deviation} or they are 
suppressed like 2HDM  by a  mixing angle. A photon 
collider offers excellent opportunities to probe these couplings.  It can be 
built from  $e^+ e^-$ or $e^- e^-$ scattering using Compton 
backscattering of lasers from $e^\pm$ beams, and it can measure 
rates for Higgs production and decays into certain final states 
with few percent level precision. Experimental studies  have been done for various photon colliders designs (TESLA, NLC, CLIC, JLC)
 of the expected precisions with which the rates $\gamma \gamma \to H \to X$ can be measured. The results show that $\gamma \gamma \to H \to \bar b b$ could be measured at about $2 \%$ level for a light Higgs boson, and the channels $W W^*$ at $5 \%$ and $Z Z^ *$ at $11\%$ \cite{gamagama}.

In this work we focus on the lightest CP-even Higgs boson $h^0$ in 
the 2HDM-III, its partial decays widths at tree level and one-loop 
level taking into account the contribution of new physics due to 
FCNC from the Yukawa couplings. We calculate the number of events 
of the $h^0$ boson as a resonance in a photon collider taking into 
account its associated background signal.

\section{The Two Higgs Doublet Model type III} 

The 2HDM includes a second Higgs doublet, and both doublets acquire 
vacuum expectation value (VEV) different from zero, given by 
\begin{equation} 
\Phi_i=\begin{pmatrix}\phi^+_i\cr\phi^0_i\end{pmatrix},\quad 
\langle\Phi_i\rangle=\begin{pmatrix}0\cr v_i/\sqrt 2\end{pmatrix}, 
\qquad i=1,2. 
\end{equation} 
The mass eigenstates have a scalar spectrum which contains two 
$CP$-even neutral Higgs bosons ($h^0$, $H^0$) coming from the 
mixing  of $\Re (\phi_i^0)$ with mixing angle $\alpha$; two charged 
Higgs bosons ($H^\pm$) which it is a mix of the would-be Goldstone bosons 
$G_W^\pm$ through the mixing angle $\tan \beta=v_2/v_1$; and one 
$CP$-odd Higgs ($A^0$) which is a mixing of the neutral would-be 
Goldstone boson $G_Z^0$. In the general 
2HDM-III the Higgs doublets can couple with the up and down 
quark sector at the same time because there is not a discrete 
symmetry and  FCNC appears at tree level. Then the Yukawa 
Lagrangian for the quarks in this model can be written as follow \cite{at-reina-soni,csy, DMR} 
\begin{equation} 
-\mathcal{L}_Y^{(III)}=\eta_{ij}^{U,0}\bar Q_{iL}^0\widetilde 
\Phi_1U_{jR}^0+\eta_{ij}^{D,0}\bar Q_{iL}^0\Phi_1D_{jR}^0+\xi_{ij}^{U,0} 
\bar Q_{iL}^0\widetilde\Phi_2U_{jR}^0+\xi_{ij}^{D,0}\bar Q_{iL}^0 
\Phi_2D_{jR}^0+\text{h.c}, 
\end{equation} 
where $\Phi_i$ are the Higgs doublets, $\eta_{ij}^0$ and $\xi_{ij}^0$ 
are non-diagonal $3\times 3$ matrices and the suffix "0" means that these 
fermion states are not mass eigenstates. From equation (2)  it is clear that 
the mass terms for the up-type or down-type sectors will depend on two Yukawa 
coupling matrices. The rotation of the quarks and leptons allow us to diagonalize 
one of the matrices but in general not both simultaneously, then one Yukawa 
coupling remains non-diagonal, leading to the FCNC at tree level. 
 
In 2HDM-III there is a global symmetry which can make a rotation 
of the Higgs doublets and fix one VEV equal to zero 
\cite{rodolfo}. In such way, $v_1=v$ and $v_2=0$, and the mixing 
parameter $\tan\beta=v_2/v_1$ can be eliminated from 
the Lagrangian. Expanding the Yukawa Lagrangian in this 
parameterization, it is found 
\begin{eqnarray} 
 -\mathcal{L}_Y^{(III)}&=&h^0\bar U\left(-\frac{s_\alpha}{v}M_U^{diag}+ 
 \frac{c_\alpha}{\sqrt 2}\xi^U\right)U+ 
 h^0\bar D\left(-\frac{s_\alpha}{v}M_D^{diag}+\frac{c_\alpha}{\sqrt 2}\xi^D\right)D 
 \nonumber \\ 
&+&H^0 \bar U \left(\frac{c_\alpha}{v}M_U^{diag}+\frac{s_\alpha}{\sqrt 
2}\xi^U\right)U+ H^0\bar 
D\left(\frac{c_\alpha}{v}M_D^{diag}+\frac{s_\alpha}{\sqrt 
2}\xi^D\right)D\nonumber \\ 
&-&\frac{i}{\sqrt 2}A^0\bar U\gamma_5\xi^UU+\frac{i}{\sqrt 
2}A^0\bar 
D\gamma_5\xi^DD-\frac{i}{v}G^0_Z\bar U\gamma_5M_U^{diag}U+\frac{i}{v}G^0_Z\bar 
D\gamma_5M_D^{diag}D\nonumber \\ 
&+&G^+_W\bar U\left\{\frac{\sqrt2 
c_\beta}{v}\left(-M_U^{diag}\mathcal{K}P_L+ 
\mathcal{K}M_D^{diag}P_R\right)\right\}D\nonumber 
\\ 
&+&H^+\bar U\left\{\mathcal{K}\xi^DP_R-\xi^U\mathcal{K}P_L\right\}D 
+\text{+ h.c.}, 
\end{eqnarray} 
where $\mathcal{K}$ is the Cabbibo-Kobayashi-Maskawa matrix and 
$s_\alpha=\sin\alpha$, $c_\alpha=\cos\alpha$. This equation shows how the FCNC processes are 
generated in this kind of model. In this case there are two matrices 
$3\times 3$ which lead to 12 new parameters if we suppose that 
these matrices are real and symmetric. 
 
In the SM the couplings $\phi f\bar f$ are proportional to $m_f$. 
Since  $\xi_{ij}$ are couplings like $\phi f_i \bar f_j$ and, if 
one wishes to avoid fine tuning, then the structure of the mass 
and the mixing hierarchy suggest that the natural parameterization 
for FCNC vertices would be the one proportional to the masses of the 
particles. In the present work we take into account the 
Cheng-Sher-Yuan (CSY) parameterization which is the geometric mean 
of the Yukawa couplings of the quark fields \cite{csy},
\begin{equation} 
\xi_{ij}\equiv\frac{\sqrt{m_im_j}}{v}\lambda_{ij}. 
\end{equation} 
This is an ansatz for the Yukawa texture matrices looking for a 
phenomenological similarity with SM couplings. Under these 
assumptions, the relative couplings are proportional to 
\begin{equation} 
R^{h^0}_{f}=-\sin\alpha+\frac{1}{\sqrt 2}\cos\alpha\lambda_{ff}, 
\end{equation}
where the parameters $\lambda_{ff}$ are those that generate a 
hierarchy between the up-type and down-type couplings. 
Bounds and 
restrictions on the $\lambda_{ij}$ for the quark sector and $\xi_{ij}$ for the leptonic sector can be found in 
literature \cite{at-reina-soni, rodolfo, bounds, htc}. Many scenarios for  the $\lambda_{ij}$ parameters have been analyzed under different phenomenological considerations. In the case of the leptonic sector, some constraints have been imposed on the $\lambda_{ij}$ using different lepton flavor violating processes and the $(g-2)_\mu$ factor. Analysis of possible detection of the Higgs boson at Tevatron and LHC using lepton flavor violating decays have been already presented \cite{diaz-cruz}. On the other hand, the quark sector has been less restricted, however three different scenarios  are discussed in reference \cite{at-reina-soni} using the available phenomenology. The first scenario discussed was assuming the whole set of $\lambda_{ij}\sim \lambda$ common to all the flavour changing couplings  but it was discarded by low energy experiments. The second one is useful to guide to the third one which have more physical relevance. The third scenario is established when the parameters related with the flavor changing vertex between first and second generation are negligible and $\lambda_{bb}$, $\lambda_{bs}$ bigger than one while $\lambda_{tt}$, $\lambda_{tc}$ smaller than one.  In table 1 we display a summary of the bounds that we  use in the present work. The leptonic element $\xi_{\mu \tau}$ has a bound coming from $(g-2)$ muon factor and it is interesting because the interval does not contain the zero like the others do. The bounds on $\lambda_{bb}$ and $\lambda_{tt}$ are gotten using the criterion of validness of perturbation theory looking at the vertex $\bar t b H^+$. On the other hand, bounds for the parameter $\lambda_{tc}$ have actually been calculated taking into account the contribution of $h^0t\bar c$ vertex at one loop level on the electroweak precision observables \cite{htc}. 
 
\begin{table}
\centering
\begin{tabular}{||c||c||c||} 
\hline\hline 
Constraint & Process & Restriction\\ 
\hline\hline 
$\xi^2_{\mu\tau}\in\left[7.62\times10^{-4};4.44\times10^{-2}\right]$ & $(g-2)_\mu$ & $m_{A^0}\to\infty$\\ 
\hline 
$\xi_{\tau\tau}\in\left[-1.8\times10^{-2};2.2\times10^{-2}\right]$ & $\tau^-\to\mu^-\gamma$ &$m_{A^0}\to\infty$\\ 
\hline 
$\xi_{\mu\mu}\in\left[-0.12;0.12\right]$ & $\tau^-\to\mu^+\mu^-\mu^-$ & $m_{A^0}\to\infty$\\ 
\hline 
$\xi_{\mu e}\in\left[-0.39;0.39\right]$ & $\tau^-\to e^-e^-\mu^+$ & $m_{A^0}\to\infty$\\ 
\hline 
$\lambda_{bb}\in\left[-100;100\right]$ & Perturbations & $v_2=0$\\ 
\hline 
$\lambda_{tt}\in\left[-\sqrt 8;\sqrt 8\right]$ & Perturbations & $v_2=0$\\ 
\hline 
$\vert \lambda_{tc} \vert  \lesssim 2.3/\cos\alpha$ & Precision test & $m_h \lesssim 170$ GeV \\ 
\hline 
\hline\hline 
\end{tabular} 
\caption{Some constraints on the flavor changing neutral parameters obtained in the literature\cite{at-reina-soni,rodolfo,bounds,htc}  and they will be taken into account in section 4.}
\end{table}

\section{$h^0$ decays in the 2HDM-III} 
 
We assume that $h^0$ is the lightest Higgs boson in the model and 
their decays  are forbidden  into another Higgs bosons.  The decay 
channels to heavy quarks and leptons are $h^0\to t\bar t, b\bar b, 
\tau^+\tau^-, \mu^+\mu^-$; and the relevant FCNC decays at tree 
level in the 2HDM type III are $h^0\to t\bar c, b\bar s, 
\mu^+\tau^-$ which in the CSY parameterization could be as high as 
the standard ones. The decay width of a scalar particle into two 
different fermions is \cite{diaz-cruz} 
\begin{eqnarray} 
\Gamma(\phi\to f_i\bar f_j)&=&\frac{N_C}{8\pi}m_\phi\left|A(\phi f_i\bar f_j)\right|^2\left(1-\left(\frac{m_i+m_j}{m_\phi}\right)^2\right)\cr&\times&\sqrt{1+\left(\frac{m_i^2-m_j^2}{m_\phi^2}\right)^2-\left(\frac{m_i^2+m_j^2}{m_\phi^2}\right)^2}, 
\end{eqnarray} 
with $N_C=1(3)$ for leptons(quarks) and $A(\phi f_if_j)$ the Feynman's rule for 
the $\phi f_if_j$ vertex.

For $h^0\to W^+W^-, Z^0Z^0$ are necessary to consider off-shell 
and the on-shell processes according to the Higgs mass. For the 
on-shell processes the widths lead to 
\begin{equation} 
\Gamma(h^0\to W^+W^-)=\frac{g^2}{64\pi}\frac{m_{h^0}^3}{m_W^2}\sin^2\alpha\left(1-4\frac{m_W^2}{m_{h^0}^2}\right)^{1/2}\left(1-4\frac{m_W^2}{m_{h^0}^2}+12\frac{m_W^4}{m_{h^0}^4}\right), 
\end{equation} 
\begin{equation} 
\Gamma(h^0\to ZZ)=\frac{g^2}{128\pi}\frac{m_{h^0}^3}{m_W^2}\sin^2\alpha\left(1-4\frac{m_Z^2}{m_{h^0}^2}\right)^{1/2}\left(1-4\frac{m_Z^2}{m_{h^0}^2}+12\frac{m_Z^4}{m_{h^0}^4}\right). 
\end{equation} 
For off-shell processes one of the gauge boson in the final state 
is a virtual gauge which can decay into two fermions. The decay 
widths in the context of the SM Higgs boson can be found in reference 
\cite{hhg}. But they have a $\sin^2\alpha$ factor, which comes from the couplings in the 2HDM. 
 
At one loop-level the considered decays are $h^0\to\gamma\gamma, gg, \gamma 
Z$. The first one, the decay width of $h^0$ into photons 
is given by 
\begin{equation} 
\Gamma(h^0\to\gamma\gamma)=\frac{\alpha^2g^2}{1024\pi^3}\frac{m^3_{h^0}}{m^2_W}\left|\sum_{i=0,1/2,1} 
N_{Ci}e_i^2F_iR_i^{h^0}\right|^2, 
\end{equation} 
where $F_i$ are functions depending of the  particle into the loop and 
$i=0,1/2,1$ correspond to scalar, fermion, and gauge boson, 
respectively, and they are defined in reference \cite{hhg}. Then, the sum is over  top and bottom quarks, $W^\pm$ gauge boson and $H^\pm$ the scalar charged Higgs boson. $R_i$ are relative couplings given by the 2HDM-III, and  we use the 
Feynmman's rule for the vertex $h^0H^+H^-$ assuming an invariant 
$CP$ potential,
\begin{equation}
h^0H^+H^-\sim -2gm_W\sin\alpha\left(\frac{1}{4\cos^2\theta_W}-1\right).
\end{equation}

The process $h^0\to gg$ is quiet similar to the last one, but it 
can only have quarks into the loop. Using again the two heaviest 
quarks the decay width is 
\begin{equation} 
\Gamma(h^0\to gg)=\frac{\alpha_s^2g^2m^3_{h^0}}{128\pi^3m_W^2} 
\left|\sum_i\tau_i\left[1+(1-\tau_i)f(\tau_i)\right] 
R_i^{h^0}\right|^2, 
\end{equation} 
with $\tau_i\equiv 4 m_i^2/m_{h^0}$ and $f(\tau_i)$ also defined 
in reference \cite{hhg}. 
 
Finally we take the process $h^0\to\gamma Z$. This loop could be 
more complicated than the others and it depends of 
fermionic, scalar and  bosonic sectors. The decay width can be written as 
\begin{equation} 
\Gamma(h^0\to\gamma Z)=\frac{\alpha^2g^2}{512\pi^3}\frac{m_{h^0}^3}{m_W^2} 
\left|A_F+A_W\right|^2\left(1-\frac{m_Z^2}{m_{h^0}^2}\right)^3, 
\end{equation} 
with the amplitudes $A_F$ and $A_W$ defined in references \cite{hhg, photonZ}. All these loop processes are sensitive 
to variations of the parameters which are model dependent.

\subsection{Different scenarios for 2HDM-III}

The Figure 1 shows the ratios $\Gamma(h^0 \to \gamma \gamma)^{2HDM}/ \Gamma(h^0 \to \gamma \gamma)^{SM}$, $\Gamma(h^0 \to g g)^{2HDM}/ \Gamma(h^0 \to gg)^{SM}$ and  $\Gamma(h^0 \to \gamma Z)^{2HDM}/ \Gamma(h^0 \to \gamma Z)^{SM}$ versus the Higgs boson mass, $m_{h^0}$, under different choices of the model parameters as it is indicated in the figure caption. The cases labeled $a$ correspond  to the asymptotic limits of the 2HDM-III to the SM values, and therefore for these cases the ratios would be equal to one. The other cases show the possible deviations from the SM. The decay widths are suppressed in the low Higgs boson mass range due to the factor $\sin \alpha$ in the $R^{h^0}_i$ couplings coming from the 2HDM.

The Figure 2 shows the branching ratios for $h^0$, when FCNC processes 
are introduced in the possible  Higgs $h^0$ decays in the framework of the  2HDM-III. The CSY 
parameterization is used and a particular case is taken into 
account when all $\lambda_{ij}$ in the quark sector are all the same order equal to 
one, but for the leptonic sector we take the values showed in table 1. Thus, it shows all the branching ratios for $h^0$ in the 2HDM-III with $1$ TeV for the charged Higgs mass  and $\alpha=(\pi/2, \pi/4,0)$. The case $\alpha=\pi/2$ and heavy charged Higgs boson  reproduces the SM  branching ratios, and it gives us the chance to check consistency;  we compare with results using computational packages as HDECAY \cite{pc}; for small values of $m_{h^0}$ the dominant decays are $b\bar b$ and $V^*V$ \cite{hhg}. For the case $\alpha=0$ the decay modes $WW$ and $ZZ$ disappear and therefore the modes $tt$ and $tc$ are the dominant modes in the region of heavy Higgs boson mass. For a light Higgs boson the dominant decay mode is $bb$, and the fraction for $h \to \gamma \gamma$ is one order of magnitude lower than the SM result because the $W$ contribution is not present into the loop and only quarks are contributing. The comparison between plots in figure 2, let us know  what happens with FCNC scenarios with respect to the SM predictions. The $\tau^+ \tau^-$ branching fraction is one order of magnitude bigger than the SM mode and for a heavy Higgs boson it is higher than the $gg$ one, contrary to the SM. On the other hand, $\gamma \gamma$ and $gg$ modes are at the same order.  The FCNC $h^0\rightarrow b\bar s$ is the same order of  $h^0\rightarrow b\bar b$ for  $m_{h^0}<$180 GeV; but in the kinematic limits for  $h^0\rightarrow t\bar c$ decay the BR($h^0\rightarrow b\bar b$)  is one order of magnitude lower than the SM.

The figure 3 shows the branching fractions taking into account  a more realistic scenario according to the phenomenological constraints for $\lambda_{ij}$ from table 1. In this case, the channels $b\bar b$ and $b\bar s$ are the leading decays in the light Higgs mass limit.  This is because 
in the range of intermediate Higgs boson mass  still dominate $b$ decays, but 
at a heavier Higgs boson 
mass,  the vector bosons reach some importance and $b$ decays  loose supremacy, opening a window to 
the $t\bar t$ channel. The leptonic $\tau$ decays are in the same order than $gg$, and always bigger 
than the other one loop decays. It is remarkable  that the FCNC channel $\mu^+\tau^-$ 
is more important than $t\bar c$, almost in one order of magnitude. The larger values for the vertex 
factors in the leptonic sector lead these processes to be relevant in the sector of about 1 decays 
through these channels each $10^4-10^5$ Higgs events. Finally we can say that the $\mu^+e^-$ is indiscernible because of the small size of the coupling. 

The figure 4 is a similar scenario to the figure 3, but the hierarchy between $\lambda_{tt(tc)}$ and $\lambda_{bb(bs)}$ is changed in one order of magnitude. The same assumption was made for the leptonic sector. In this case, the branching ratios for the $h^0$ into fermions without flavour changing are only one order of magnitude bigger than the similar ones with flavour changing.

In the 2HDM-III the relation between two branching ratios can be written as
\begin{equation}
r_{ff/ff'}= \frac{B(h^0 \to f \bar f)}{B(h^0 \to f \bar f')} \propto \left \vert \frac{- \sqrt{2} s_\alpha + c_\alpha \lambda_{ff}}{\lambda_{ff'} c_\alpha}   \right \vert^2 \, ,
\label{razon}
\end{equation}
taking off kinematic factors. When $\lambda_{ff} \simeq \lambda_{ff'}$ but bigger than one the ratio $r_{ff/ff'}\simeq 1$, it means that the branching ratios are at the same order of magnitude for any value of $\alpha$ different of $\alpha= \pi/2$. It is the scenario presented in figure 4. However if $\lambda_{ff} \simeq \lambda_{ff'}$ but smaller than one the ratio $r_{ff/ff'}$ is proportional to $(\lambda_{ff} c_\alpha)^{-1}$ which implies that the branching ratio $B(h^0 \to f \bar f)$ is going to be bigger than the branching ratio $B(h^0 \to f \bar f')$. It is precisely the situation showed in figure 3. Furthermore, in figure 3 when $\alpha$ goes from $3 \pi/8$ to zero the ratio $r_{tt/tc}$ goes from $10^4 -10^5$ to $10^2$.


 
\section{$h^0$ production and detection at photon collider} 
 
A photon collider provides a scenario to look for new physics 
\cite{pcol}. The collider $\gamma\gamma$, $\gamma e$ are based on 
Compton backscattering of laser light off the high energy 
electrons of linear collider. Most important photon colliders 
which are under construction are TESLA \cite{TDR}, 
NLC\cite{nlc}, JLC \cite{jlc} and CLIC \cite{clic}. These 
colliders have a huge importance, because typical cross sections 
of interesting processes in $\gamma\gamma$ collisions, as charged 
pair production, are higher than those in $e^+e^-$ collisions by 
about one order of magnitude, so the number of events in 
$\gamma\gamma$ collisions will be larger than $e^+e^-$ collisions. 
Besides, these photons will have a high degree of circular 
polarization, allowing different $J^{PC}$ states more than 
$e^+e^-$ collider. 
 
The production of neutral scalars at photon colliders is mediated 
by one loop-level processes, and it is interesting to see that 
$\phi^0\gamma\gamma$ vertex is sensitive to small variations of 
particle couplings into the loop. In particular, different 
scenarios of 2HDM-III parameters will produce a huge change in the 
$h^0\gamma\gamma$ vertex. This sensitive vertex will allow the possibility of distinguish the scalar boson from different models \cite{maria}. In photon colliders the number of events of $h^0$ produced as a 
resonance of fermions in the final state,  $\gamma\gamma\to 
h^0\to f_i\bar f_j$, is given by \cite{haberhelicity}
\begin{eqnarray} 
N(\gamma\gamma\to h^0\to f_i\bar f_j)&=&8\pi F(y_{h^0})\mathcal{L}_{e^+e^-}(1+\langle\lambda\lambda' 
\rangle_{y_{h^0}})\frac{\Gamma(\gamma\gamma\to h^0)B(h^0\to f_i\bar f_j)}{m_{h^0}^2 
E_{e^+e^-}}\cr&\times&\arctan\left[\frac{\Gamma_{res}}{\Gamma_{h^0}}\right], 
\end{eqnarray} 
where, $\lambda\lambda'$ is the product of the helicities of the two colliding photons, 
$F(y_{h^0})\mathcal{L}_{e^+e^-}$ is the differential luminosity for the $\gamma\gamma$ 
collider and the resolution width is defined by $\Gamma_{res}\equiv\max\left\{\Gamma_{exp}, 
\Gamma_{h^0}\right\}$ with $\Gamma_{exp}$ the experimental width allowed by the detector.
 
The background signal is given by the process $\gamma\gamma\to 
f\bar f$, and the number of events are \cite{haberhelicity} 
\begin{equation} 
N(\gamma\gamma\to f\bar f)=\frac{\Gamma_{res}}{E_{e^+e^-}}F(y_{h^0})\mathcal{L}_{e^+e^-}\sigma_{\gamma\gamma\to f\bar f}(m_{h^0}^2,z_0), 
\end{equation} 
where 
\begin{eqnarray} 
\sigma_{\gamma\gamma\to f\bar f}(s,z_0)&=&\frac{4\pi\alpha^2e^4_fN_C}{s}\left\{-\beta z_0\left[1+\frac{(1-\beta^2)^2}{1-\beta^2z_0^2}\right]+\frac{3-\beta^4}{2}\ln\frac{1+\beta z_0}{1-\beta z_0}\right.\cr &+&\left.\lambda\lambda'\beta z_0\left[1+\frac{2(1-\beta^2)}{1-\beta^2z_0^2}-\frac{1}{\beta z_0}\ln\frac{1+\beta z_0}{1-\beta z_0}\right]\right\}, 
\end{eqnarray} 
with $\beta\equiv\left(1-4m_f^2/s\right)^{1/2}$ and 
$z_0=\cos\theta_0$ the maximum sweep detector angle. It is 
interesting to note that the cross section is evaluated in the 
$h^0$ resonance and, worthwhile that the background is proportional to $e^4_f$.

TESLA (TeV Electron Superconducting Linear Accelerator), will work 
with a solid angle resolution about $z_0=0.85$, an average 
polarization $\langle\lambda\lambda'\rangle_{h^0}=0.8$, a 
luminosity d $L_{\gamma\gamma}(z_0\geq0.8)=1.15\times10^{34}$ 
cm$^{-2}$ s$^{-1}$ in a year, a CM energy $E_{e^+e^-}$=500 GeV and 
an experimental resolution about 5 GeV \cite{TDR}. 

In figure 5 using the same parameters as figure 4, we can see the number of events of the Higgs signal (s) according 
to the 2HDM-III and its corresponding background signal (b). The $b\bar b$ channel is in fact the best channel for $h^0$ 
decays, and for light Higgs boson mass. It is always above its own 
background signal. It is different form the SM case where 
only  in a small region  near to $120-170$ GeV, the signal is bigger 
than the background signal\cite{haberhelicity}. On the 
other hand, the background signal for heavy lepton pair production 
is bigger than those decays into quarks.  This result was 
expected, because photons couple through the electromagnetic 
charge, and  leptons has integer charge, therefore its background 
will not have any suppression factor, as already happens with 
quarks. The $h^0$ signal from leptonic decays is undetectable, 
owing to have one Higgs boson which decays in leptons each 
$10^4$ charged pair $\tau^+\tau^-$ produced.  Now, the background signal for $t\bar t$ is always bigger at least one order of magnitude, however the detection of this $10\%$ of signal is strongly dependent of the top decays.

Let us 
examine what happens with FCNC processes in the quark sector, 
$t\bar c$ and $b\bar s$. We start with $t\bar c$ process. Charm 
quark could be detected through its jet, but top quark decay in 
$bW^+$. Bottom quark would be tagged and $W$ boson could suffer 
leptonic decays $l^+\bar\nu$ or can generate  a quark pair. 
 The channel with $W$ leptonic decays can be easily 
reconstructed by lost energy and information from the 
electromagnetic calorimeter, so the most promising signal would be $jet+b-tagged+l+ \not E$. For the hadronic decay of the $W$ boson, a background possibility is  $b+b + 2\, jets$, this case will have  
difficulties for $h^0$ detection because the $t\bar c$ signal is 
smaller than $b\bar b$ background signal. For the $b\bar s$ channel, 
we would have a $jet$ and a $b-tagged$. In general, these FCNC channels will 
have not background, and we will have some statistics 
for $h^0$ detection. A deeper analysis with noise from NLO QCD 
correction could be found in reference \cite{roeck}. It is important 
remark that the analysis for FCNC process in the quark sector, is quite similar for the others photon colliders.

CLICHE (Compact Linear Collider Higgs Experiment) \cite{clic}, developed on CLIC 1 at low energy will work with $\langle\lambda\lambda'\rangle_{h^0}=0.94$, a geometrical position of the detector at $z_0=0.85$, the luminosity $L_{\gamma\gamma}(z_0\geq0.8)=4.7\times 10^{34}$ cm$^{-2}$ s$^{-1}$, a CM energy $E_{e^+e^-}=150$ GeV and an experimental resolution given by 3.3 GeV. 

The figure 5 shows the behavior of $h^0$ at CLIC photon collider. It will be working at a lower energy than TESLA and therefore some QED and QCD phenomena that can appear in the laser vanish, increasing the number of events. CLIC could only create too light $h^0$ as far as 150 GeV, through $b\bar b$ channel, which is the most indicated channel for light scalar detection. CLIC will produce near to $10^6$ events of $h^0$ in a year, by using $\gamma\gamma\to h^0\to b\bar b$ signal, while the background will be the $5\%$ of the signal, being this a great scenario to measure the light $h^0$ properties. It is obvious that there are no kinematic conditions for top quark production. Now, the FCNC process $b\bar s$ will be  bigger than $b\bar b$ background signal, then as we already mentioned  will be this channel easily reconstructed, leaving a huge window for FCNC detection at CLIC. For leptonic decays, again the background affect enormously the $h^0$ signal. The $\tau^+\tau^-$ $h^0$ signal is around $0.01\%$ of the background, which makes the $h^0$ Higgs boson undetectable through leptonic decays. Something similar happens with $\mu^+\tau^-$ process. 
 
 
Finally we will analyze the $h^0$ production and detection at NLC (Next Linear Collider) \cite{nlc}. This collider will work with a polarization average $\langle\lambda\lambda'\rangle_{h^0}=0.79$, $L_{\gamma\gamma}(z\geq 0.8)=3.4\times 10^{34}$ cm$^{-2}$ s$^{-1}$ for a CM energy $E_{e^+e^-}=450$ GeV and the experimental width of the detector near to 13.1 GeV. 

The Figure 5 shows the number of $h^0$ expected in NLC through fermionic decays and its noise signal. If $m_{h^0}$ is below $\sim 275$ GeV then the Higgs boson should be detected and its couplings determinate in NLC through $b\bar b$ decay, due to this signal larger than the background signal.  Same as TESLA, the background of $t\bar t$ channel is bigger than its $h^0$ signal, this obstruct the detection in pair top production. The leptonic FCNC process do not reach to appear in the figure, and the charged lepton pair production with $h^0$ in the resonance loose importance because of the $\tau^+\tau^-$ and $\mu^+\mu^-$ background signal. Again, the interesting FCNC channel are given by the quark sector. The $b\bar s$ show a production of 5000 $h^0$ events near to the top resonance, getting too close to $b\bar b$ background signal. In this region the FCNC process will be easily detected, but in the rest of the spectrum the detection will be strongly determinate by the b-tagging within the vertex detector.

 
\section{Conclusion} 
 
The addition of a second Higgs doublet, as an extension of the SM, lead to different new physics scenarios. One of them is the 2HDM-III where discrete symmetries are not imposed and it generates a general Yukawa Lagrangian with FCNC at tree level.  This fact makes the 2HDM-III a very complex model and there are not relations between  the scalar spectrum as already happen in 2HDM-II. In the Yukawa sector there are twelve new parameters which generate FCNC or modify the SM $h^0$ couplings. 

We first study the decay widths of the lightest $CP$-even Higgs boson $h^0$ at tree level and at one loop level. The new physics effects can be tested into the loops for different values of the FCNC parameters, and hence different 2HDM-III scenarios can be proposed.  In particular, the case for $\alpha=\pi/2$ and a heavy charged Higgs boson reproduces the SM predictions for the branchings ratios. In the other cases the one loop widths are lower than SM. For high Higgs boson mass range the 2HDM has parameter scenarios where $\Gamma(h^0\rightarrow \gamma\gamma)$ is bigger than SM width. 
 
In figure 2 we consider one scenario where all $\lambda_{ij}$ in the quark sector are equal to one \cite{at-reina-soni}. In this case the new FCNC at tree level are of the same order of the SM ones. However, when $\alpha=\pi/2$ the new physics is decoupled and the SM's branchings are reproduced for a heavy charged Higgs boson.  
 
In the other figures we take into account more realistic scenarios for the $\lambda_{ij}$ coming from the 2HDM contributions to different phenomenology processes. The branching ratios of the one loop decays are one or two orders of magnitude lower than the SM due to the new channels $h^0\rightarrow b\bar{s}$ and $h^0\rightarrow t\bar{c}$ which are at the same order of $h^0\rightarrow b\bar{b}$ and $h^0\rightarrow t\bar{t}$, respectively.  A very particular scenario is when  $\alpha=0$. In this scenario $WWh^0$ and $ZZh^0$ couplings are equal to zero,  and  $b\bar{b}$ and $t\bar{t}$ are the most important channels for light and heavy Higgs boson, respectively. But the FCNC at tree level $b\bar{s}$ and $t\bar{c}$ decay modes are close to them.  Now, in  $\gamma\gamma$ and $gg$ branching ratios the contributions into the loops are only coming from fermions because the gauge sector is decoupled into the loops in the $h^0 \to \gamma\gamma$ decay. 
 
On the other hand, for the 2HDM-III we write down the expression (\ref{razon}) where it is saying that for big values of $\lambda_{ff} \simeq \lambda_{ff'}$, the decays involving  flavour change are at the same order that those without flavour change. Otherwise for small values of $\lambda_{ij}$, the flavour changing channels are too small. All these means that the detection of fermionic flavour changes at tree level is not enough to test the 2HDM-III. However if the branching ratios or the number of events of $h^0$ produced satisfy a relationship like equation (\ref{razon}), that could be the first indication that 2HDM-III is important to reveal the presence of new physics. 

Finally, we have set  up one SM-like scenario but introducing FCNC processes and we analyze the $h^0$ detection and production at photon colliders such as TESLA, CLIC and NLC. We use fermionic decays and the cross sections evaluated in the $h^0$ resonance. We calculate the background signals for $h^0$ decays, using the helicity formalism in order to do a naive analysis of the possible noise within the detector. The background signals are proportional to $e_f^4$, and then lepton channels have bigger noise than quarks channels. The signal for the $h^0$ decays into a pair of quarks are one order of magnitude higher than the similar decays with flavor changing, however the second one does not have any background signal. This is an interesting scenario to produce and detect Higgs bosons and a possibility to test new physics beyond the SM.    
 
We would like to thank Colciencias and DIB-UN for financial support.

\newpage

\begin{figure}

\includegraphics[angle=0]{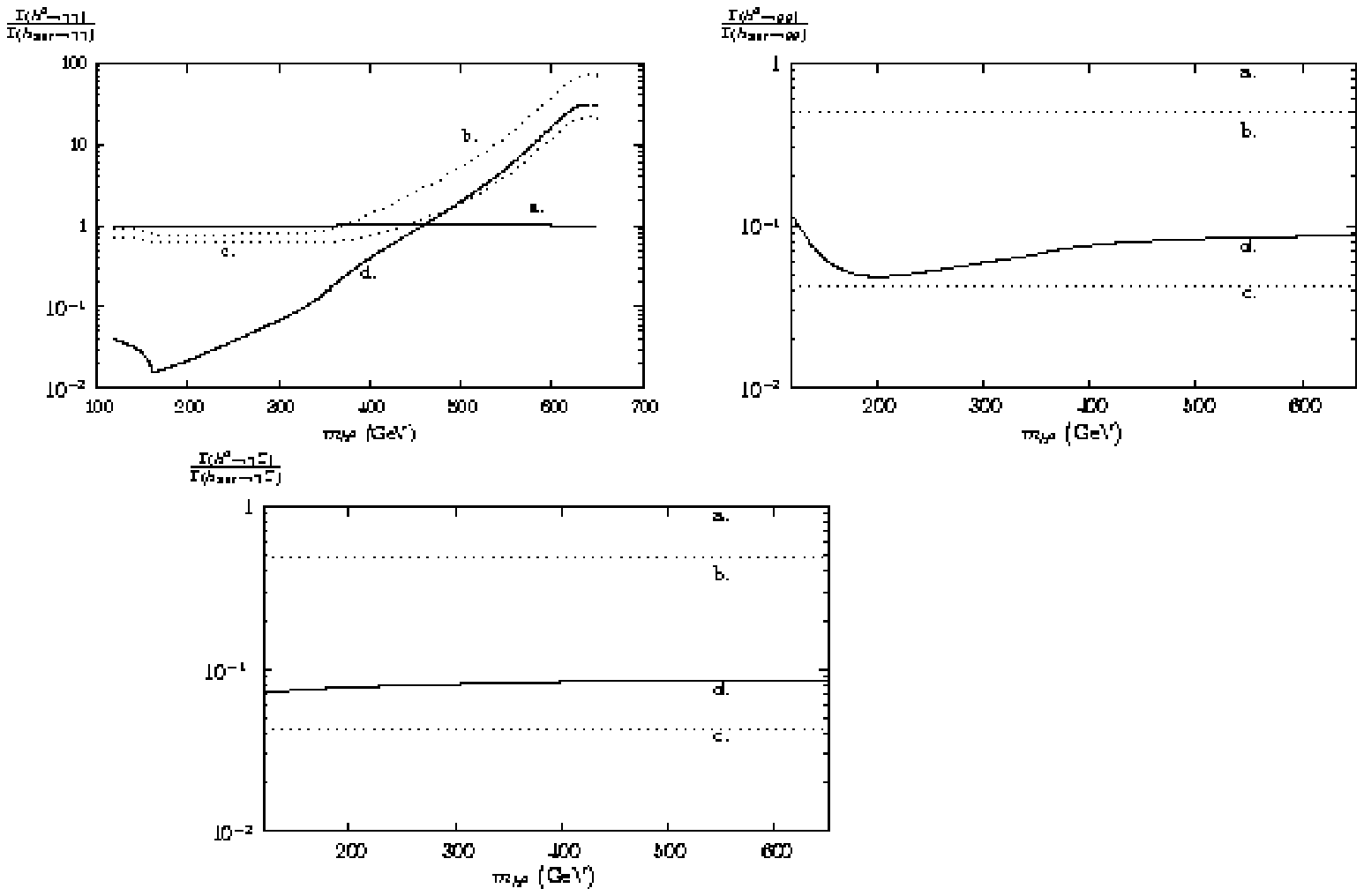}

\caption{Sensitivity of $\frac{\Gamma(h^0\to\gamma\gamma)}{\Gamma(h_{SM}\to\gamma\gamma)}$, $\frac{\Gamma(h^0\to gg)}{\Gamma(h_{SM}\to gg)}$ and $\frac{\Gamma(h^0\to\gamma Z)}{\Gamma(h_{SM}\to\gamma Z)}$ using different values of 2HDM-III parameters. a) $\alpha=\pi/2$, $m_{H^\pm}=1$ TeV; b) $\alpha=\pi/4$, $m_{H^\pm}=500$ GeV, $\lambda_{tt}=1$, $\lambda_{bb}=10$; c) $\alpha=\pi/4$, $m_{H^\pm}=500$ GeV, $\lambda_{tt}=1$, $\lambda_{bb}=1$; d) $\alpha=0$, $m_{H^\pm}=500$ GeV, $\lambda_{tt}=1$, $\lambda_{bb}=1$.} 
\end{figure}

\bigskip
\pagebreak
\newpage

\begin{figure}
\includegraphics[angle=0]{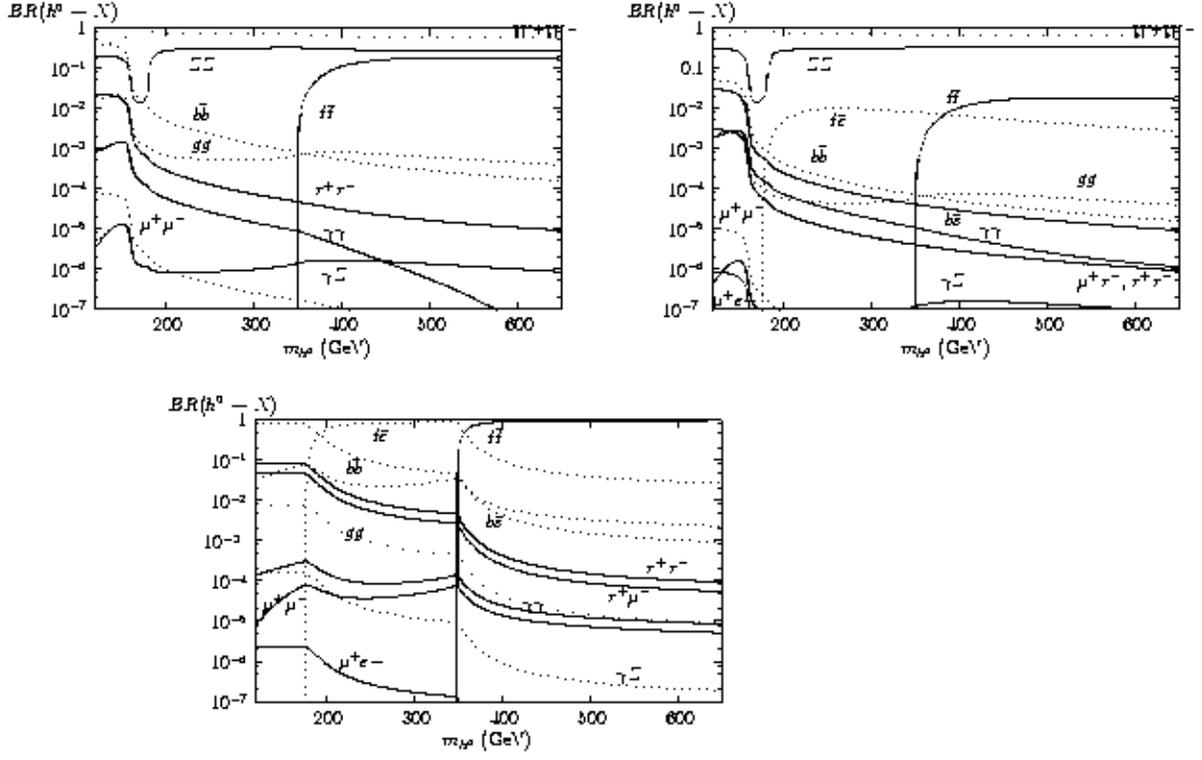}
\caption{$h^0$ Branching ratios in the 2HDM with $\lambda_{ij}=1$ and $\alpha=\pi/2,\pi/4,0$}
\end{figure}

\begin{figure}
\includegraphics[angle=0]{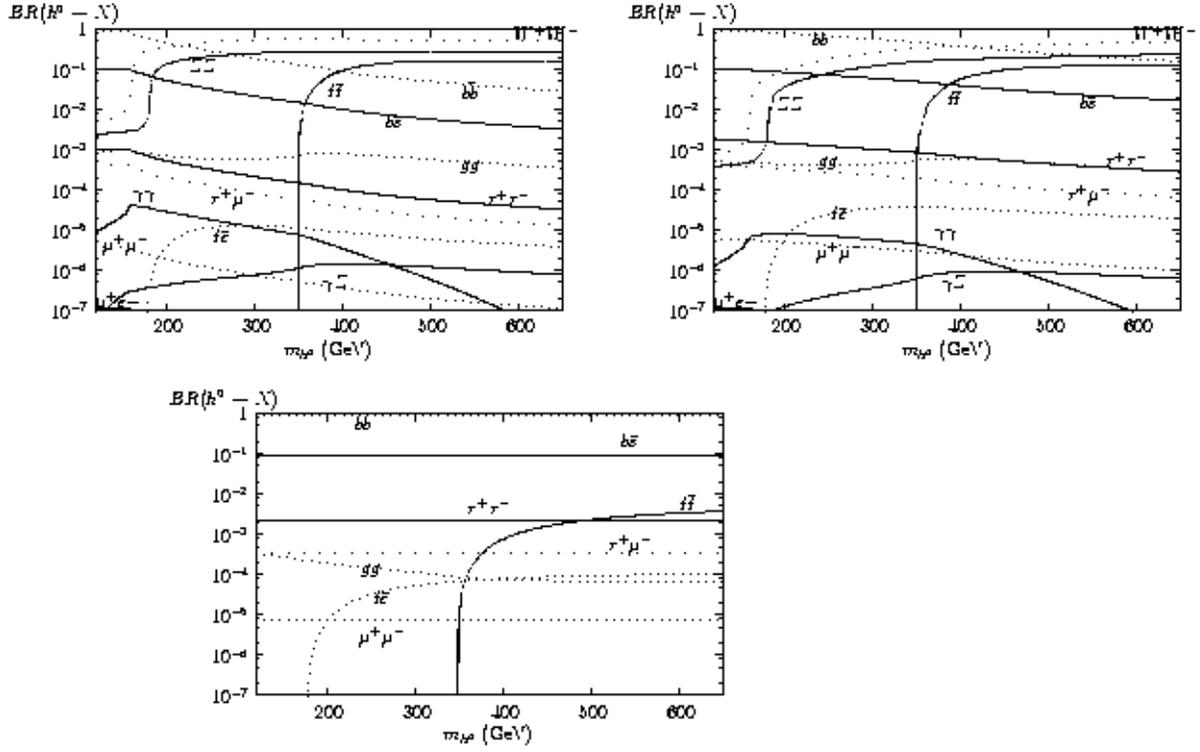}
\caption{Branching ratios for $\lambda_{tc}=\lambda_{tt}=0.1$,  $\lambda_{bs}=\lambda_{bb}=50$,  $\lambda_{\mu\mu}=\lambda_{\tau\tau}=\lambda_{\mu\tau}=\lambda_{e\mu}=10$, and $\alpha=3\pi/8,\pi/4,0$.}
\end{figure}

\begin{figure}
\includegraphics[angle=0]{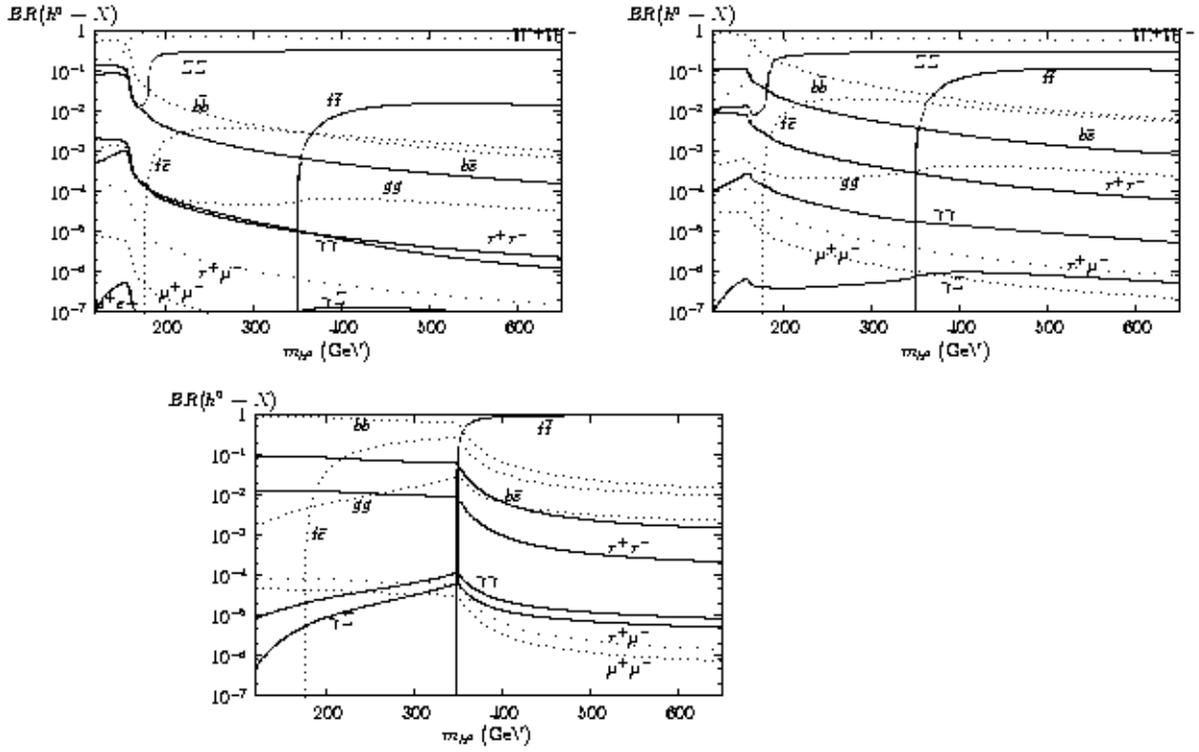}
\caption{ Branching ratios for $\lambda_{tc}=1.5$,$\lambda_{tt}=2.5$,  $\lambda_{bs}=\lambda_{bb}=10$,  $\lambda_{\mu\mu}=\lambda_{\tau\tau}=5$, $\lambda_{\mu\tau}=1$, and $\alpha=3\pi/8,\pi/4,0$.}
\end{figure}

\newpage

\begin{figure} 
\includegraphics[angle=0]{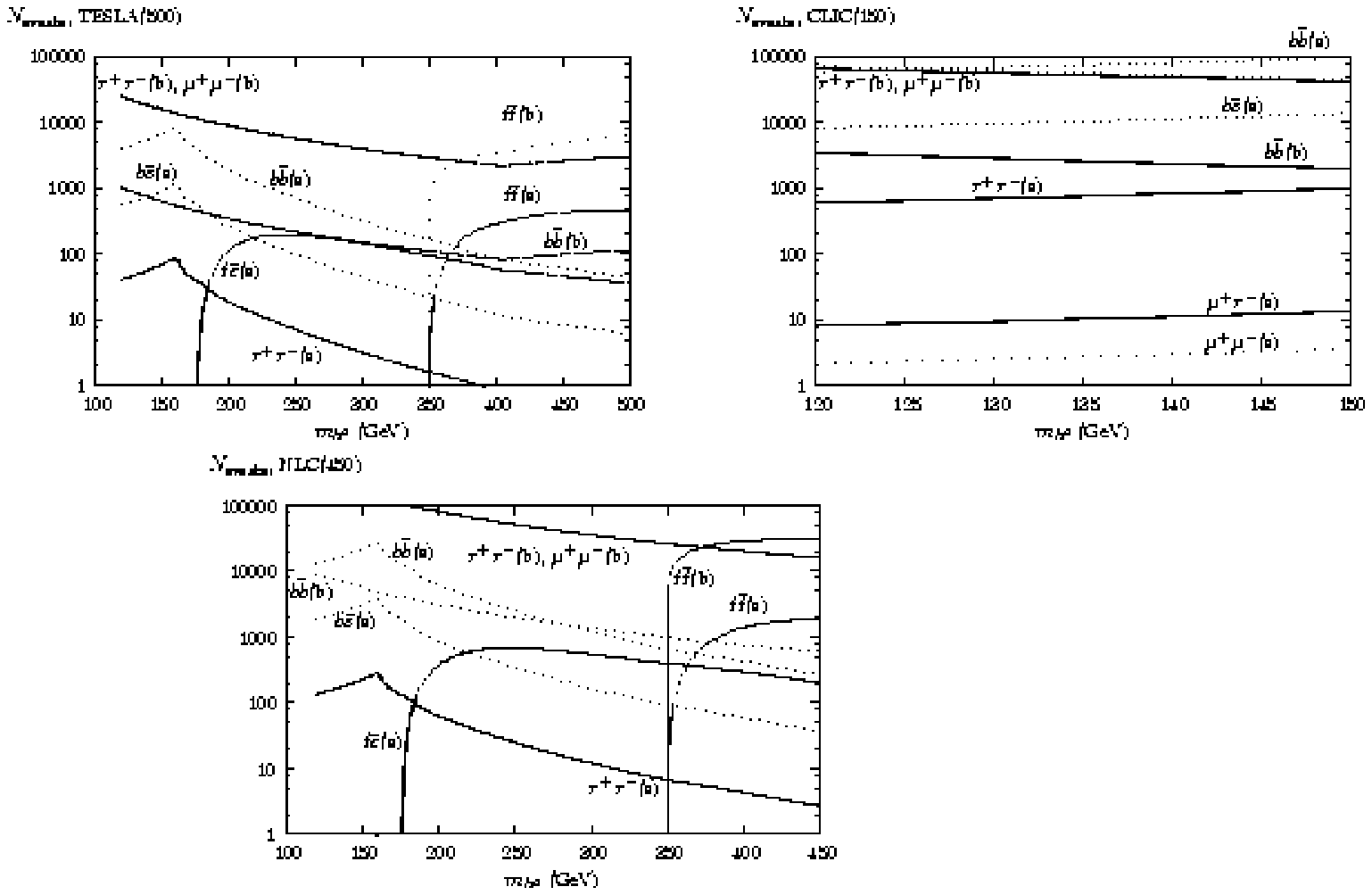}
\caption{Number of events of $h^0$ and the corresponding background in TESLA(500), CLIC(150) and NLC(450)} 
\end{figure} 


\begin{thebibliography}{99}
 
\bibitem{SM} S. Weinberg. \textit{Phys. Rev. Lett. }\textbf{21}, 1264 
(1967).S. Salam, in \textit{Elementary Particle Theory}, Ed. N. 
Southolm (1968). S. L. Glashow, \textit{Nucl. Phys.} \textbf{22}, 
579 (1961). 

\bibitem{susy}
B.~C.~Allanach {\it et al.},
in {\it Proc. of the APS/DPF/DPB Summer Study on the Future of Particle Physics (Snowmass 2001) } ed. N.~Graf,
Eur.\ Phys.\ J.\ C {\bf 25}, 113 (2002)
[eConf {\bf C010630}, P125 (2001)]
[arXiv:hep-ph/0202233];H.~E.~Haber,
Nucl.\ Phys.\ Proc.\ Suppl.\  {\bf 101}, 217 (2001)
[arXiv:hep-ph/0103095].

\bibitem{extra}N.~Arkani-Hamed, S.~Dimopoulos and G.~R.~Dvali,
Phys.\ Lett.\ B {\bf 429}, 263 (1998)
[arXiv:hep-ph/9803315];I.~Antoniadis, N.~Arkani-Hamed, S.~Dimopoulos and G.~R.~Dvali,
Phys.\ Lett.\ B {\bf 436}, 257 (1998)
[arXiv:hep-ph/9804398];.~Randall and R.~Sundrum,
Phys.\ Rev.\ Lett.\  {\bf 83}, 3370 (1999)
[arXiv:hep-ph/9905221];L.~Randall and R.~Sundrum,
Phys.\ Rev.\ Lett.\  {\bf 83}, 4690 (1999)
[arXiv:hep-th/9906064].
 
 
 \bibitem{strong}W.~A.~Bardeen, C.~T.~Hill and M.~Lindner,
Phys.\ Rev.\ D {\bf 41}, 1647 (1990);H.~Georgi, D.~B.~Kaplan and P.~Galison,
Phys.\ Lett.\ B {\bf 143}, 152 (1984);D.~B.~Kaplan, H.~Georgi and S.~Dimopoulos,
Phys.\ Lett.\ B {\bf 136}, 187 (1984).

\bibitem{little}N.~Arkani-Hamed, A.~G.~Cohen and H.~Georgi,
Phys.\ Lett.\ B {\bf 513}, 232 (2001)
[arXiv:hep-ph/0105239];N.~Arkani-Hamed, A.~G.~Cohen, T.~Gregoire and J.~G.~Wacker,
JHEP {\bf 0208}, 020 (2002)
[arXiv:hep-ph/0202089];N.~Arkani-Hamed, A.~G.~Cohen, E.~Katz, A.~E.~Nelson, T.~Gregoire and J.~G.~Wacker,
JHEP {\bf 0208}, 021 (2002)
[arXiv:hep-ph/0206020];
N.~Arkani-Hamed, A.~G.~Cohen, E.~Katz and A.~E.~Nelson,
JHEP {\bf 0207}, 034 (2002)
[arXiv:hep-ph/0206021]; T.~Han, H.~E.~Logan, B.~McElrath and L.~T.~Wang,
Phys.\ Rev.\ D {\bf 67}, 095004 (2003)
[arXiv:hep-ph/0301040].

\bibitem{osc} R. N. Mohapatra, [arXiv:hep-ph/0211252] (2002). J. Bernabeu, 
[arXiv:hep-ph/0012312] (2000). J. W. F. Valle, 
[arXiv:hep-ph/0410103] (2004). 

\bibitem{GW} S. L. Glashow and S. Weinberg,  \textit{Phys. Rev.} 
\textbf{D15}, 1958 (1977). 

\bibitem{hep-ex0412015}OPAL Collaboration (The LEP Electroweak Working Group,{\it et.al}),
[arXiv:hep-ex/0412015].
 
\bibitem{precision} W. J. Marciano, Invited 
talks at SSI2004 SLAC Summer Institute, [arXiv:hep-ph/0411179] 
(2004); U.~Baur {\it et al.}  [The Snowmass Working Group on Precision Electroweak  Measurements],
in {\it Proc. of the APS/DPF/DPB Summer Study on the Future of Particle Physics (Snowmass 2001) } ed. N.~Graf,
eConf {\bf C010630}, P1WG1 (2001)
[arXiv:hep-ph/0202001].

\bibitem{LHC}M.~Duhrssen, S.~Heinemeyer, H.~Logan, D.~Rainwater, G.~Weiglein and D.~Zeppenfeld,
Phys.\ Rev.\ D {\bf 70}, 113009 (2004)
[arXiv:hep-ph/0406323];M.~Carena, S.~Heinemeyer, C.~E.~M.~Wagner and G.~Weiglein,
Eur.\ Phys.\ J.\ C {\bf 26}, 601 (2003)
[arXiv:hep-ph/0202167];K.~A.~Assamagan {\it et al.}  [Higgs Working Group Collaboration],
arXiv:hep-ph/0406152. 

\bibitem{linear}T.~Abe {\it et al.}  [American Linear Collider Working Group],
in {\it Proc. of the APS/DPF/DPB Summer Study on the Future of Particle Physics (Snowmass 2001) } ed. N.~Graf,
arXiv:hep-ex/0106056; J.~A.~Aguilar-Saavedra {\it et al.}  [ECFA/DESY LC Physics Working Group],
arXiv:hep-ph/0106315; K.~Abe {\it et al.}  [ACFA Linear Collider Working Group],
arXiv:hep-ph/0109166.


\bibitem{deviation}H.~E.~Logan,
Phys.\ Rev.\ D {\bf 70}, 115003 (2004)
[arXiv:hep-ph/0405072];G.~A.~Gonzalez-Sprinberg, R.~Martinez and J.~A.~Rodriguez,
arXiv:hep-ph/0406178, to appear PRD

\bibitem{gamagama}D.~Asner {\it et al.},
Eur.\ Phys.\ J.\ C {\bf 28}, 27 (2003)
[arXiv:hep-ex/0111056];D.~Asner {\it et al.},
arXiv:hep-ph/0308103;T.~Ohgaki, T.~Takahashi and I.~Watanabe,
Phys.\ Rev.\ D {\bf 56}, 1723 (1997)
[arXiv:hep-ph/9703301];D.~M.~Asner, J.~B.~Gronberg and J.~F.~Gunion,
Phys.\ Rev.\ D {\bf 67}, 035009 (2003)
[arXiv:hep-ph/0110320; G.~Jikia and S.~Soldner-Rembold,
Nucl.\ Phys.\ Proc.\ Suppl.\  {\bf 82}, 373 (2000)
[arXiv:hep-ph/9910366]; P.~Niezurawski, A.~F.~Zarnecki and M.~Krawczyk,
Acta Phys.\ Polon.\ B {\bf 34}, 177 (2003)
[arXiv:hep-ph/0208234].
 
\bibitem{at-reina-soni}D.~Atwood, L.~Reina and A.~Soni,
Phys.\ Rev.\ D {\bf 55}, 3156 (1997)
[arXiv:hep-ph/9609279]. 

\bibitem{csy} T. P. Cheng and M. Sher. \textit{Phys. Rev. 
}\textbf{D35}, 3484 (1987). M. Sher and Y. Yuan. \textit{Phys. 
Rev. }\textbf{D44}, 1461 (1991). M. Sher, [arXiv:hep-ph/9809590] 
(1998). 
 
\bibitem{DMR} R. A. Diaz, R. Martinez and 
J-A. Rodriguez, \textit{Phys. Rev. }\textbf{D63}, 095007 (2001), 
[arXiv:hep-ph/0010149]. 

\bibitem{rodolfo} R. A. Diaz, PhD Thesis, Universidad Nacional de Colombia, 
[arXiv:hep-ph/0212237] (2002) and references there in. 

\bibitem{bounds} R. A. Diaz, R. Martinez and C. E. 
Sandoval, [arXiv:hep-ph/0311201] (2003). R. A. Diaz, R. Martinez 
and C. E. Sandoval, [arXiv:hep-ph/0406265] (2004). R. A. Diaz, R. 
Martinez and J-A. Rodriguez, [arXiv:hep-ph/0103050] (2001). R. 
Martinez, J-A. Rodriguez and M. Rozo,  \textit{Phys. Rev. 
}\textbf{D68}, 035001 (2003), [arXiv:hep-ph/0212236].

\bibitem{htc}F.~Larios, R.~Martinez and M.~A.~Perez,
arXiv:hep-ph/0412222.

\bibitem{diaz-cruz}J.~L.~Diaz-Cruz, R.~Noriega-Papaqui and A.~Rosado,
Phys.\ Rev.\ D {\bf 69}, 095002 (2004)
[arXiv:hep-ph/0401194]; J.~L.~Diaz-Cruz and J.~J.~Toscano,
Phys.\ Rev.\ D {\bf 62}, 116005 (2000)
[arXiv:hep-ph/9910233].

\bibitem{hhg} J. F. Gunion, H. E. 
Haber, G. Kane and S. Dawson. \textit{The Higgs Hunter's Guide}. 
Addison Wesley (1990). 

\bibitem{photonZ} R. Martinez, M. A. Perez,  \textit{Nucl. Phys. 
}\textbf{B347}, 105 (1990).
 
\bibitem{pc} A. Pukhov, \textit{et.al.}, [arXiv:hep-ph/9908288] (1999). A. Djouadi, J. Kalinowski and M. Spira,  \textit{Comp. Phys Commun. 
}\textbf{108}, 56 (1998), [arXiv:hep-ph/9704448]. \bibitem{pcol} 
I. Ginzburg,\textit{Nucl. Instrum. Meth. }\textbf{A472}, 121 
(2001), [arXiv:hep-ph/0101029]. A. De Roeck, J. Ellis and F. 
Gianotti, [arXiv:hep-ex/0112004] (2001). S. J. Brodsky, 
\textit{Int. J. Mod. Phys. } \textbf{A18}, 2871 (2003), 
[arXiv:hep-ph/0204197]. 

 
\bibitem{TDR} B. Badeleck \textit{et.al.}, 
[arXiv:hep-ex/0108012] (2001). V. Telnov, [arXiv:hep-ph/0010033] 
(2001). E. Boos, \textit{et. al.}, \textit{Nucl. Instrum. Meth.} 
\textbf{A472}, 100 (2001), [arXiv:hep-ph/0103090]. 
 
\bibitem{nlc}M. M. Velazco \textit{et. al.}, [arXiv:hep-ex/0111055] (2001). 
 
\bibitem{jlc}I. Watanabe, \textit{et.al.} KEK-REP-97-17 (1997). 
 
\bibitem{clic}D. Asner, \textit{et.al.}, CERN-TH/2001-235, 
[arXiv:hep-ph/0111056] (2001). 

 
\bibitem{maria}P.~Niezurawski, A.~F.~Zarnecki and M.~Krawczyk,
arXiv:hep-ph/0403138;
I.~F.~Ginzburg, M.~Krawczyk and P.~Osland,
arXiv:hep-ph/0101331;
M. Krawczyk, 
[arXiv:hep-ph/0307314] (2003). O. Yavas and A. K. Ciftci, 
[arXiv:hep-ex/0004013] (2000). V. Strakhovenko, X. Artru, R. 
Chehab and M. Chevallier, [arXiv:hep-ph/0409178] (2004). 
 
 
\bibitem{haberhelicity} H. E. Haber, [arXiv:hep-ph/9405376] (1994). 
J. F. Gunion and H. E. Haber, \textit{Phys. Rev.  }\textbf{D48}, 
5109 (1993).


 
\bibitem{roeck}A. De Roeck, [arXiv:hep-ph/0311138] (2003) and 
references cited. 

 
 


\end{thebibliography}
\end{document}